\documentclass[journal=nalefd,manuscript=letter,layout=traditionnal]{achemso}
\setkeys{acs}{articletitle = true}
\usepackage{xcolor}
\usepackage{bm}
\usepackage{amsmath}

\usepackage{caption}
\usepackage{mfirstuc}
\usepackage[utf8]{inputenc}

\author{Juan Carlos Estrada Salda{\~n}a}
\affiliation{Univ. Grenoble Alpes, INAC-PHELIQS, F-38000 Grenoble, France and CEA, INAC-PHELIQS, F-38000 Grenoble, France}
\altaffiliation{Present address: Center for Quantum Devices, Niels Bohr Institute, University of Copenhagen, 2100 Copenhagen, Denmark}
\author{Yann-Michel Niquet}
\affiliation{Univ. Grenoble Alpes, INAC-MEM, F-38000 Grenoble, France and CEA, INAC-MEM, F-38000 Grenoble, France}
\author{Jean-Pierre Cleuziou}
\author{Eduardo J. H. Lee}
\affiliation{Univ. Grenoble Alpes, INAC-PHELIQS, F-38000 Grenoble, France and CEA, INAC-PHELIQS, F-38000 Grenoble, France}
\altaffiliation{Present address: Condensed Matter Physics Center (IFIMAC), Universidad Aut\'{o}noma de Madrid, 28049 Madrid, Spain.}
\author{Diana Car}
\affiliation{Technische Universiteit Eindhoven, P.O. Box 513, 5600 MB Eindhoven}
\author{S{\'e}bastien R. Plissard}
\affiliation{CNRS, LAAS-CNRS, Université de Toulouse, 31400 Toulouse, France}
\author{Erik P. A. M. Bakkers}
\affiliation{Technische Universiteit Eindhoven, P.O. Box 513, 5600 MB Eindhoven}
\author{Silvano De Franceschi}
\affiliation{Univ. Grenoble Alpes, INAC-PHELIQS, F-38000 Grenoble, France and CEA, INAC-PHELIQS, F-38000 Grenoble, France}
\email{silvano.defranceschi@cea.fr}

\title{\capitalisewords{Split-Channel ballistic transport\space in\space an InSb nanowire}}

\keywords{Conductance quantization, One-dimensional transport, Nanowire, Band structure, Helical liquid, Majorana fermions}

\begin{document}

\begin{abstract}

We report an experimental study of one-dimensional (1D) electronic transport in an InSb semiconducting nanowire. Three bottom gates are used to locally deplete the nanowire creating a ballistic quantum point contact with only a few conducting channels. In a magnetic field, the Zeeman splitting of the corresponding 1D subbands is revealed by the emergence of conductance plateaus at multiples of $e^2$/h, yet we find a quantized conductance pattern largely dependent on the configuration of voltages applied to the bottom gates. In particular, we can make the first plateau disappear leaving a first conductance step of 2$e^2/h$, which is indicative of a remarkable two-fold subband degeneracy that can persist up to several Tesla. For certain gate voltage settings, we also observe the presence of discrete resonant states producing conductance features that can resemble those expected from the opening of a helical gap in the subband structure. We explain our experimental findings through the formation of two spatially separated 1D conduction channels.  
\end{abstract}

\maketitle

One-dimensional (1D) nanowires made of small band-gap semiconductors, such as InSb or InAs, have been identified as a promising material system for the realization of topological superconductivity \cite{stanescu2013majorana,lutchyn2017realizing}. This exotic state of matter, which is characterized by the emergence of zero-energy Majorana quasiparticles localized at the nanowire edges, is expected to occur only when some key requirements are simultaneously met \cite{Oreg2010,lutchyn2010majorana}. 
Above all, the nanowire has to be 1D and at most moderately disordered \cite{brouwer2011probability,lobos2012interplay,degottardi2013majorana}.

The linear conductance of a clean, ballistic 1D nanowire is expected to be quantized. At zero magnetic field, $B$, time-reversal symmetry leads to doubly degenerate subbands yielding conductance plateaus at multiples of 2$e^2/h$, where $e$ is the electron charge and $h$ the Planck constant \cite{van1988quantized}. This two-fold degeneracy gets lifted at finite $B$ leading to $e^2/h$ conductance steps.

While the low-temperature transport properties of semiconductor nanowires have been studied for more than fifteen years \cite{li2006nanowire}, the achievement of ballistic 1D conduction has been challenging due to the presence of disorder of various origin \cite{voisin2014few,schroer2010correlating,wallentin2011electron,salfi2011probing,holloway2013trapped,weis2014quantum}. Following some initial experimental signatures in Ge/Si core/shell nanowires \cite{lu2005one}, further experimental evidence of conductance quantization was more recently reported using either InAs \cite{ford2012observation,abay2013quantized,vigneau2014magnetotransport,heedt2016ballistic,heedt2017signatures,gooth2017ballistic,gooth2017connects} or InSb nanowires \cite{doi:10.1021/nl3035256,Kammhuber2016,zhang2016ballistic,fadaly2017observation}. In these studies, conductance plateaus were observed by sweeping the voltage of a \textit{single} gate creating a tunable electrostatic potential barrier in the nanowire. Here we investigate conductance quantization in an InSb nanowire simultaneously gated by three closely spaced electrodes. This increased complexity results in an enhanced tunability of the local conduction-band profile giving access to unexpected properties of the 1D electronic subbands. 

\begin{figure}
	\includegraphics[width=0.8\linewidth]{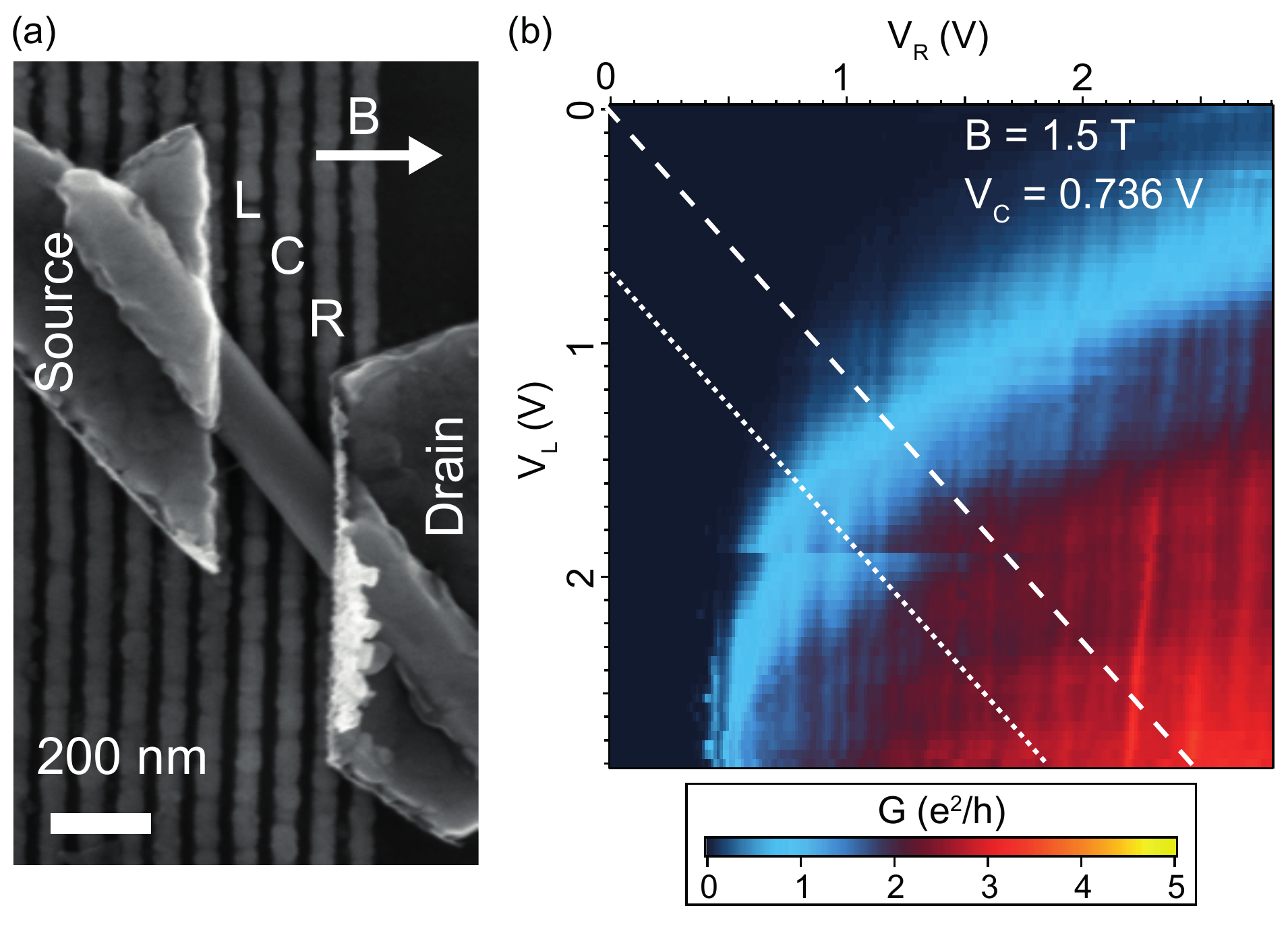}
\caption{(a) Scanning electron micrograph of the device. The magnetic field was applied in the plane of the substrate and at an angle of 58$^{\circ}$ relative to the nanowire axis. (b) Linear conductance, $G$, as function of the two lateral gate voltages, $V_L$ and $V_R$, at fixed voltage $V_C$ on the central gate. The dotted (dashed) white line indicates the gate sweeping trajectory for the data of Fig. \ref{fig:Fig2} (Fig. \ref{fig:Fig4}).}
\label{fig:Fig1}
\end{figure}

\begin{figure}
	\includegraphics[width=0.65\linewidth]{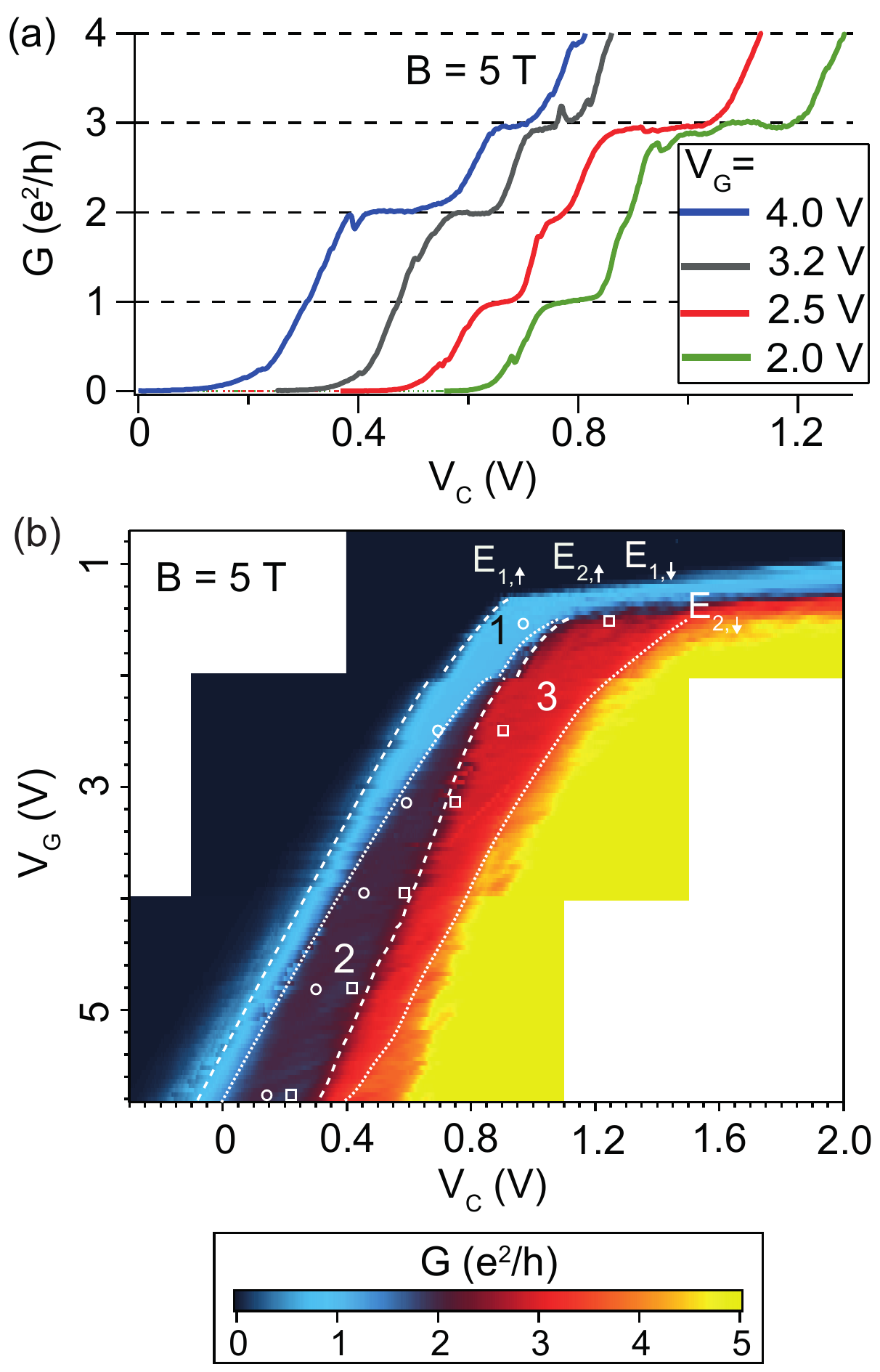}
\caption{(a) Influence of $V_G$ on the linear conductance G($V_C$) at $B = 5 $ T. The $e^2/h$ plateau is missing in the leftmost (blue) curve and the $2e^2/h$ plateau has almost entirely vanished in the rightmost (green) curve. (b) Linear conductance maps at $B = 5 $ T as a function of the central ($V_C$) and the $V_G$ gate voltages. The bottom energies of the $\uparrow$ and $\downarrow$ spin components of the two first subbands are indicated by dashed (1$^{st}$ subband) and dotted (2$^{nd}$ subband) lines. Open circles and squares indicate the barycenter of the two subbands, which provide a measure of the subband spacing $\Delta$E.}
\label{fig:Fig2}
\end{figure}

The studied device, shown in Fig. \ref{fig:Fig1}a, was fabricated from of a 160-nm-diameter InSb nanowire.  In an optical setup equipped with a micro-manipulator, the nanowire was picked from its growth substrate and deposited on a bed of bottom gates covered by 10 nm of hafnium oxide.  Source and drain contacts were defined by e-beam lithography followed by Ar sputtering and in-situ e-beam evaporation of 120-nm-thick Al. Three local bottom gates, labeled as L (left), C (central) and R (right), were used to deplete the nanowire channel. Magneto-transport measurements were performed in a He-3 refrigerator with a base temperature of 260 mK.

Figure \ref{fig:Fig1}b shows a color plot of the linear conductance, $G$, as a function of voltages $V_L$ and $V_{R}$ applied to gates $L$ and $R$, respectively. Gate $C$ is simultaneously set to $V_C = 0.736$ V, and an in-plane magnetic field $B=1.5$ T is applied as indicated in Fig. \ref{fig:Fig1}a.   
This measurement shows that $V_L$ and $V_{R}$ have a balanced effect with comparable capacitive couplings reflecting the nominally identical widths of the corresponding gates. In the following, we shall present data sets where gates L and R are swept together, along diagonal trajectories in the  ($V_L$,$V_{R}$) plane, as indicated by the white lines in Fig. \ref{fig:Fig1}b. The underlying idea is that sweeping along these lines from top-left to bottom-right corresponds to changing the longitudinal potential landscape from a double-humped to a single-humped camel back shape. At the transition, the conduction band is expected to present a single, relatively long and approximately flat potential barrier whose height can be adjusted by varying $V_C$.

The first data set is shown in Fig. \ref{fig:Fig2}, where gate voltages $V_L$ and $V_{R}$ are swept along the dotted line in Fig. \ref{fig:Fig1}b. In our notation, each gate-voltage sweep is parametrized by $V_G$, such that $V_L = V_G$ and $V_{R}=\alpha_{R/L}(V_{L}-V_{L0})$, with $V_{L0}$ = 0.74 V and $\alpha_{R/L}$ = 0.875.   
A set of $G(V_C)$ traces taken at $B = 5$ T at four different values of $V_G$ is displayed in Fig. \ref{fig:Fig2}a. The traces show plateaus of quantized conductance at multiples of $e^2/h$. Remarkably, the gate width of the plateaus varies with $V_G$. In particular, the $e^2/h$ and the 3$e^2/h$ plateaus smoothly shrink as $V_G$ is increased, whereas the 2$e^2/h$ plateau simultaneously broadens.
At such a high $B$, the large Zeeman splitting, $E_Z$, expected in InSb should lift the two-fold subband degeneracy and give rise to a first conductance step of  $e^2/h$, as indeed observed earlier \cite{doi:10.1021/nl3035256}. Instead, at $V_G$ = 4 V (blue trace) the first $e^2/h$ plateau has completely vanished leaving a $2e^2/h$ conductance step. This behavior would be compatible with the hypothesis of a lowest energy subband with largely suppressed g-factor, resulting in a quasi-two-fold degeneracy. In fact, we provide an alternative interpretation as discussed further below. 

A first intuition of the underlying scenario can be acquired from the two-dimensional color plot of G($V_C$,$V_G$) shown in Fig. \ref{fig:Fig2}b. Here the color scale has been adjusted to emphasize the plateaus at $e^2/h$, $2e^2/h$, $3e^2/h$ , and $4e^2/h$. Dashed (dotted) lines denote the conductance steps associated with the onset of conduction through the two pseudo-spin components, $\uparrow$ and $\downarrow$, of the first (second) spin-split subband. These lines correspond to aligning the spin-resolved subband edges with the Fermi energies of the leads (which coincide in the linear regime, i.e. for no dc bias-voltage between source and drain). We denote the energies of the subband edges as 
$E_{n,\uparrow}$ and $E_{n,\downarrow}$, where $n=1,2$ is the orbital quantization number.

This assignment leads us to the following considerations. 
The two pseudo-spin components of the second subband exhibit an almost $V_G$-independent spacing, which may suggest a roughly constant Zeeman splitting. This is not the case for the first subband though. In fact, the relationship between $V_C$ spacing and Zeeman splitting is not so straightforward. From a careful analysis of finite-bias measurements (i.e. differential conductance as a function of source-drain bias voltage, $V_{sd}$, and $V_C$), we find that the observed $V_G$ dependence of the subband splittings originates mostly from variations in the lever-arm parameter relating $V_C$ to subband energy. This analysis is presented in the Supplementary Information. Interestingly, it reveals that the two spin-split subbands have significantly distinct lever arm parameters, denoting different capacitive couplings to gate C. Upon increasing $V_G$ from 2 to 4 V, the lever-arm parameter of the first spin-split subband, $\alpha_1$, varies from $0.062 \pm 0.010$ to $0.037 \pm 0.008$ eV/V. The one for the second spin-split subband, $\alpha_2$, is less affected by $V_G$, varying from $0.040 \pm 0.006$ to $0.029 \pm 0.005$ eV/V, the variation occurring mainly between $V_G = 2$ and $2.5$ V. 
Within the experimental uncertainty,  the two spin-split subbands exhibit the same Zeeman energies and, correspondingly, the same g-factors, i.e. $g_1 \approx g_2 \approx 46$ (absolute value, 20\% uncertainty). These g-factors are consistent with those obtained from tight-binding calculations (see Supplementary Information).

With regard to the orbital subband splitting, defined as $\Delta E = (E_{2,\uparrow} + E_{2,\downarrow})/2 - (E_{1,\uparrow} + E_{1,\downarrow})/2)$, we find that it decreases with $V_G$, going from $\Delta E = 12.5 \pm 3.5$ meV at $V_G=2$ V to $\Delta E = 3 \pm 2$ meV at $V_G=4$ V. This evolution is apparent from the relative $V_C$ spacing between the barycenters of the first and second spin-split subbands, indicated by open circles and squares, respectively (each of these symbols is located half-way between the $V_C$ positions of the $\uparrow$ and $\downarrow$ subband components).

Given the large g-factors, an even moderate $B$ can induce crossings between subbands with opposite spin. For low $V_G$, where $\Delta E$ is maximal, the $\uparrow$ component of the second subband is expected to cross the $\downarrow$ component of the first subband for $B \sim 4.5$ T. As a result, at $B = 5$ T  $E_{2,\uparrow}$ lies below $E_{1,\downarrow}$ for every $V_G$ value. For relatively high $V_G$ values, however, $E_{2,\uparrow}$ approaches $E_{1,\uparrow}$ closely due to the largely reduced 
$\Delta E$, resulting in a virtual degeneracy and a first conductance step of $2e^2/h$, instead of $e^2/h$. Interestingly, this virtual degeneracy, occurring between parallel spin states, persists over a large $V_G$ range, extending up to the largest value explored (5.8 V). From the discussion of Fig. \ref{fig:Fig2}a, we conclude that tuning the potential landscape through $V_G$ has a significant effect on the orbital subband splitting. 

\begin{figure}
	\includegraphics[width=0.9\linewidth]{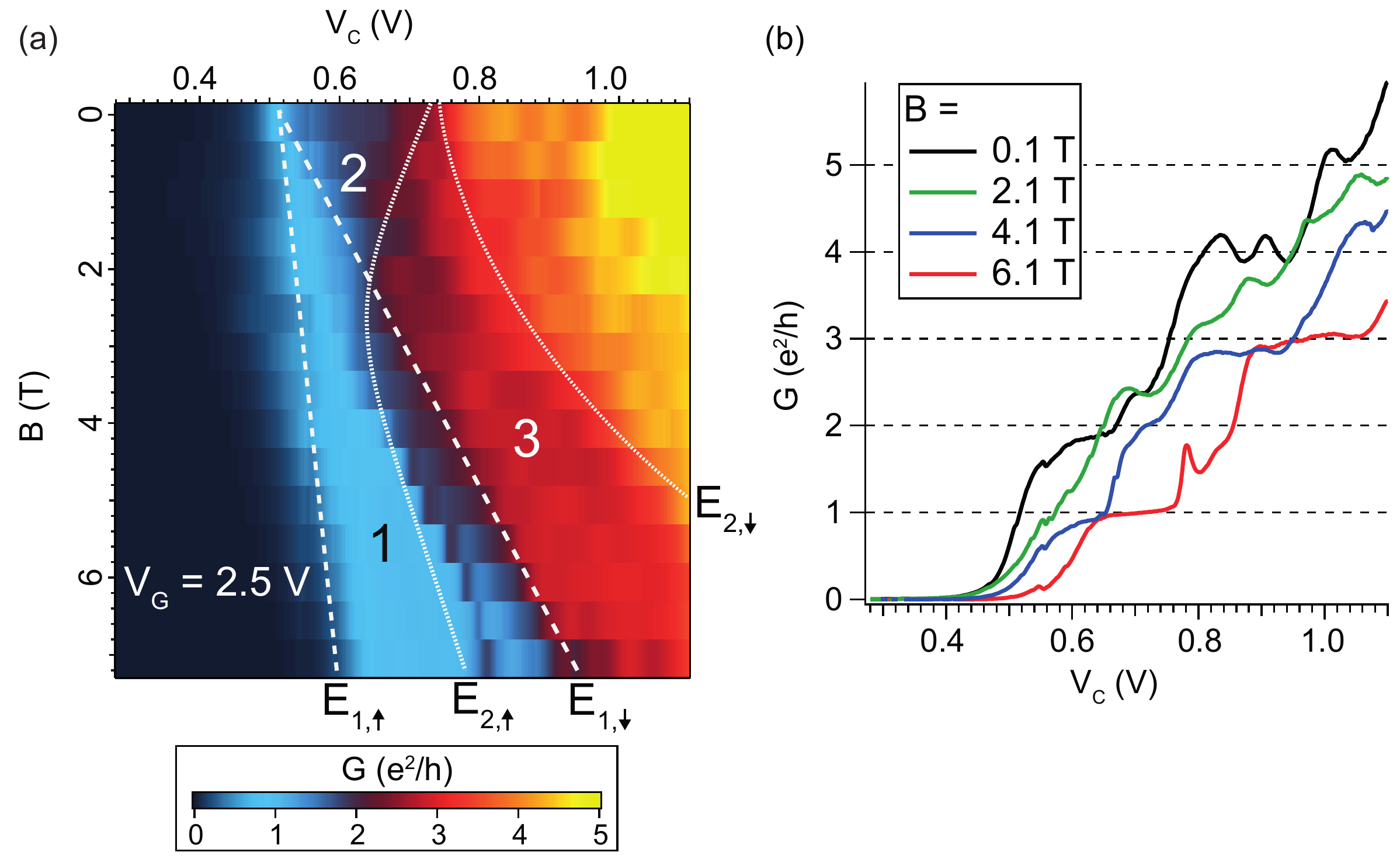}
\caption{(a) Color plot of G($V_C$,$B$) at $V_G$ = 2.5 V. The dashed and dotted lines have the same meaning as in Fig. \ref{fig:Fig2}b, and were drawn after imposing the same g-factors as in that case and allowing for adjustable $B$-induced orbital shifts. A crossing between $E_{2,\uparrow}$ and $E_{1,\downarrow}$ occurs at $\approx 2$ T. (b) G($V_C$) line cuts from the map in (a) at different fields. At B = 6.1 T (red curve), a dip of conductance can be observed in correspondence of the $2e^2/h$ plateau around $V_C = 0.8$ V.}
\label{fig:Fig3}
\end{figure}

We now turn to the $B$-dependence of the 1D subbands. We investigate that for a fixed value of $V_G$.
Figure \ref{fig:Fig3}a shows a map of $G$ as a function of $V_C$ and $B$ for $V_G = 2.5$ V. In the limit of vanishing $B$, $G$ exhibits distinguishable plateaus at $2e^2/h$ and $4e^2/h$, consistently with the expected two-fold Kramers degeneracy of the first two subbands. This is clearly visible in a $G(V_C)$ line-cut at $B=0.1$ T as shown in Fig. \ref{fig:Fig3}b (here, the small applied $B$ is necessary to suppress superconductivity in the Al-based contacts). Noteworthy, the plateaus are not as sharply quantized as at high $B$. The superimposed conductance modulation is due to a stronger back-scattering at $B=0.1$ T.  

Increasing $B$ results in a rapid splitting of the subband spin degeneracies.   
Using the same notation as in Fig. \ref{fig:Fig2}b, the onset of conduction through the different spin-resolved subbands has been indicated by dashed and dotted lines. In order to draw such lines, we have allowed for adjustable $B$-induced orbital shifts, and we have assumed $B$-independent g-factors equal to those measured at $B=5$ T. This constrain was imposed to ensure consistency with the data of Fig. \ref{fig:Fig2}b and to enable drawing lines also where the conductance steps are not clearly visible, such as the one associated with $E_{2,\uparrow}$ in the field range $B < 4$ T. 

The drawn lines follow fairly well the $B$-evolution of the spin-resolved conductance steps. According to our analysis, the subband edges $E_{2,\uparrow}$ and $E_{1,\downarrow}$ cross each other at $B \approx 2$ T. Around this $B$, however, the conductance plateaus cannot be discriminated, as apparent also from the $G(V_C)$ trace at $B=2.1$ T in Fig. \ref{fig:Fig3}b). This crossing between $E_{2,\uparrow}$ and $E_{1,\downarrow}$ implies that at higher $B$, where conductance quantization emerges again, the $\uparrow$-spin channel of the second subband opens up before the $\downarrow$-spin channel of the first one. As a result, the three plateaus at $e^2/h$, $2e^2/h$, and $3e^2/h$, clearly visible in the 4.1 T $G(V_C)$ trace in \ref{fig:Fig3}b or in the red trace of Fig. \ref{fig:Fig2}a (at $B=5$ T), correspond to the onset of conduction through the $\uparrow$-spin channel of the first subband, the  $\uparrow$-spin channel of the second subband, and the $\downarrow$-spin channel of the first subband, respectively.

\begin{figure}
	\includegraphics[width=0.82\linewidth]{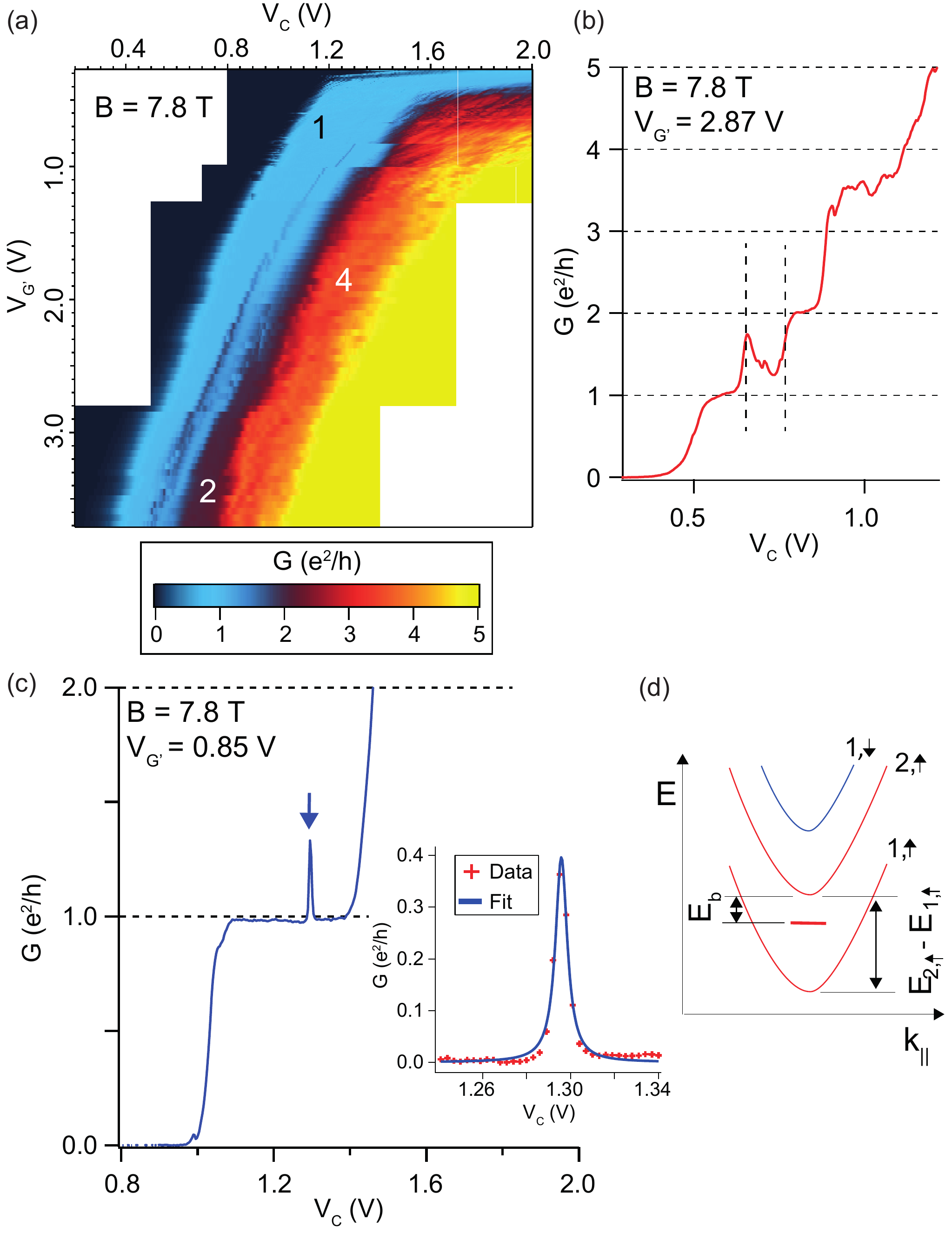}
\caption{(a) $G(V_C,V_{G'})$ maps at B = 7.8 T; $V_{G'}$ parametrizes a sweep in ($V_{L}$,$V_{R}$) along the dashed line of Fig. \ref{fig:Fig1}b. A sharp resonance on top of the $e^2/h$ plateau, related to the second spin-polarized subband, evolves into a broad peak when $V_{G'}$ is raised. (b) Line cut from (a) at $V_{G'} = 2.87$ V, when the resonance is wide. The two vertical dashed lines indicate the energy extension 
($E_b \approx 5.5$ meV) of the dip of conductance just after the resonance. (c) Line cut from (a) at $V_{G'} = 0.85$ V, when the resonance (blue arrow) becomes sharp. Inset: Lorentzian fit of the resonance after removal of the background conductance. (d) Schematic band diagram showing the localized state with binding energy $E_b$ relative to the second spin-up subband.}
\label{fig:Fig4}
\end{figure}

For $B > 6$T, a pronounced conductance suppression develops in correspondence of the second plateau. Phenomenologically, this behavior resembles very much the one recently \cite{kammhuber2017conductance} observed by Kammhuber \textit{et al.} and interpreted as a signature of a so-called ``helical gap", namely an energy interval with an effectively spinless 1D mode \cite{Oreg2010}. When the electrochemical potential lies within this gap, $G$ is supposed to drop from $2e^2/h$ to $e^2/h$, due to the lifted spin degeneracy. In combination with induced superconductivity, this spinless condition is essential to the development of Majorana-fermion states. Therefore, the direct experimental observation of helical 1D modes represents an important milestone in the field of topological superconductivity and Majorana fermions. From the $V_C$ extension of the conductance dip, approximately 110 mV, we would deduce an energy gap of $E_b \approx 5.5$ meV. This value is comparable to the one found by Kammhuber \textit{et al.}. Yet it exceeds by two orders of magnitude the spin-orbit energy given by tight binding calculations (see Supplementary Information), and the one measured in InSb nanowire quantum dots\cite{nadj2012spectroscopy}. 
Here, however, by exploiting the tunability offered by the multiple gate geometry of our device, we show that the observed feature has in fact a different nature.

Figure \ref{fig:Fig4}a shows a $G(V_C,V_{G'})$ plot similar to the one of Fig. \ref{fig:Fig2}a, where $V_{G'}$ has replaced $V_{G}$ to parametrize the simultaneous sweeping of $V_L$ and $V_R$ along the white dashed line in Fig. \ref{fig:Fig1}b, slightly shifted from the previous sweeping line (precisely: $V_L = V_{G'}$ and $V_{R}=\alpha_{R/L}(V_{L}-V_{L0})$, with $V_{L0}$ = 0 and $\alpha_{R/L}$ = 0.875). This plot is a collage of multiple measurements all taken at $B = 7.8$ T. A $G(V_C)$ line cut at $V_{G'} = 2.87$ V, corresponding to values of $V_L$ and $V_R$ close to those of Fig. \ref{fig:Fig3}, is shown in Fig. \ref{fig:Fig4}b. In correspondence of the $2e^2/h$ plateau we find the conductance suppression observed in the 6.1 T trace of Fig. \ref{fig:Fig3}b. By analyzing its evolution as a function of $V_{G'}$ we can make some important observations. Upon decreasing $V_{G'}$,  the conductance dip widens and deepens down to $e^2/h$. On its right hand side, the dip is followed by a well-defined $2e^2/h$ plateau. On its left hand side, however, the conductance approaches $2e^2/h$ for large $V_G'$, but it progressively decreases with $V_G'$ acquiring a sharp peak structure, which is characteristic of a tunnel resonance through a discrete state, as shown in the line cut at $V_{G'} = 0.85$ V in Fig. \ref{fig:Fig4}c.

The inset of Fig. \ref{fig:Fig4}c shows a Lorentzian fit of this peak after subtraction of the $e^2$/h background. The resonance width corresponds to an energy broadening 
$\Gamma$ = 0.29 $\pm$ 0.05 meV. This value largely exceeds the one expected from the temperature-induced smearing of the Fermi distribution function (i.e.  3.5 $k_B T = 0.08$ meV, where $k_B$ is the Boltzmann constant). Therefore, $\Gamma$ is dominated by the life-time broadening of the discrete state due to tunnel coupling with the leads. The peak width  increases with $V_G'$, while its position follows the $2e^2/h$ plateau keeping a constant spacing from the corresponding conductance step.

From the line shape of the conductance peak, we conclude that it arises from resonant tunneling through a bound state originating from the second spin-polarized subband. This localized state appears to be largely decoupled from the first subband (fitting to a Fano-shape line yields an only marginally better result). The dip width is determined by the binding energy of the localized state relative to the second spin-split subband. This interpretation rules out the original hypothesis of a conductance dip associated with a helical gap. An additional line cut from the top-most region of Fig. \ref{fig:Fig4}a (displayed in the Supplementary Information) reveals the presence of a conductance peak also before the onset of the first conductance plateau at $e^2/h$. This implies that, depending on the electrostatic potential landscape, a localized state can also emerge from the first spin-polarized subband.

\begin{figure}
	\includegraphics[width=0.85\linewidth]{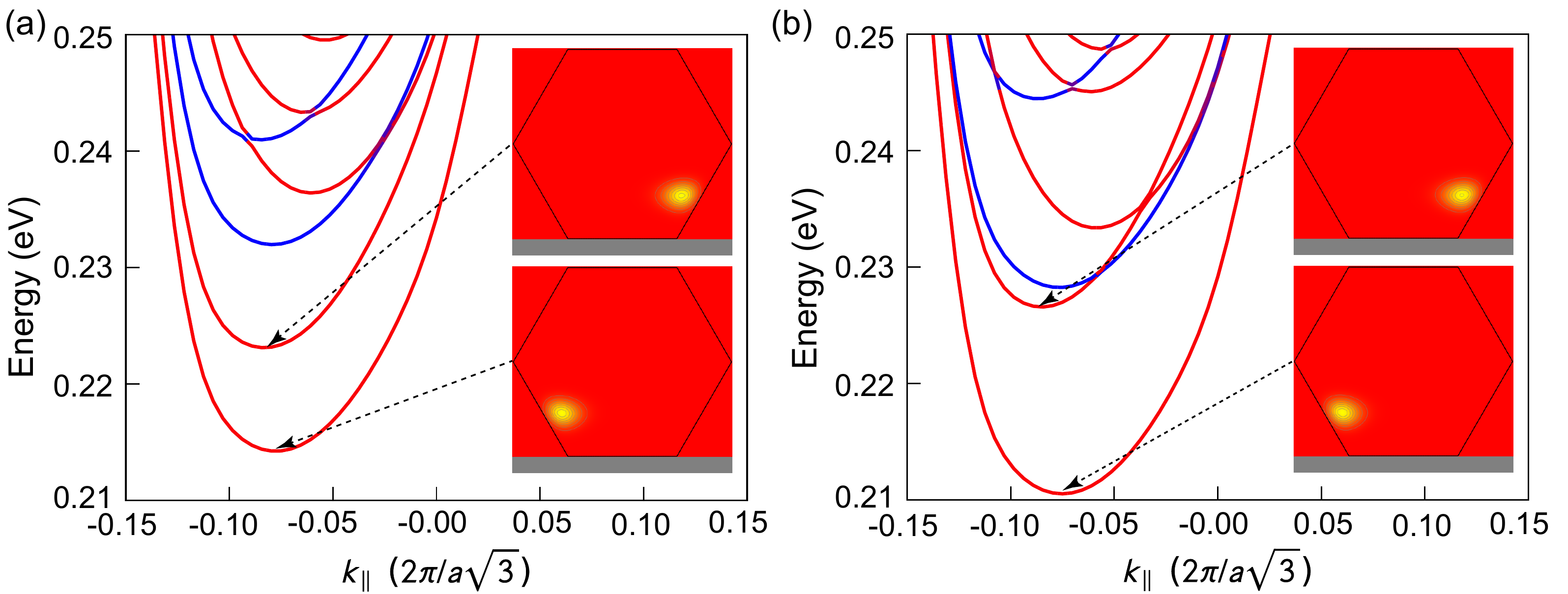}
\caption{(a,b) Tight-binding band structure of an hexagonal InSb nanowire (diameter 160 nm) at magnetic field $B=8$ T aligned as in the experiment. The calculation also accounts for the effect of gate-induced electric fields and surface charges (see main text and Supplementary Information for details). $k_\parallel$ is the longitudinal wave vector and $a$ is the lattice parameter of InSb. The red (blue) curves correspond to spin-up (spin-down) subbands. The squared wave functions of the lowest two subbands are plotted in the insets. They are strongly localized, extending over just $\sim 10$ nm. The vertical confinement is due to the applied B, while the horizontal one is due to the in-plane electric field $E_y$ from the gates and surface charges. The degeneracy between the left and right channels is lifted by a small $E_y$ = 0.1 meV/nm in (a) and $E_y$ = 0.18 meV/nm in (b), yielding band structures approximately consistent with the experimental data of Fig. \ref{fig:Fig4}b and \ref{fig:Fig4}c, respectively.}
\label{fig:Fig5}
\end{figure}

The absence of quantum mechanical coupling between the resonant bound state and the first spin-polarized subband in Fig. \ref{fig:Fig4} suggests that the corresponding wave functions are spatially separated. To investigate this possibility, we have performed tight-binding calculations with an in-plane magnetic field aligned as in the experiment. The simulations take also into account the effect of the electric fields arising from gate biasing and surface charges (see Supplementary Information for a detailed discussion of these simulations). Figure \ref{fig:Fig5} shows results at magnetic field $B=8$ T, for a distribution of negative charges at the InSb/HfO$_2$ interface. The energy of the lowest energy subbands is plotted as a function of the longitudinal wave-vector $k_\parallel$. As observed in the experiment, the first two subbands have the same spin-up polarization (we used red for up spins and blue for down spins). The squared wave functions of these two subbands are shown in the insets. They are strongly confined, vertically by the magnetic field, and laterally by the electric field from the surface charges. The first two subbands actually correspond to two side channels, which are split here by $\sim 10$ meV, comparable to the measured splitting. This splitting results from and is proportional to a small lateral electric field ($\sim 0.1$ mV/nm), which we have deliberately introduced to account for a plausible slight asymmetry between the left and right of the nanowire (due, e.g., to an imbalance of surface charges or to the fact that the gates are not running perpendicular to the nanowire). These simulations show that electric and magnetic confinement can easily lead to the formation of distinct (uncoupled) channels in such nanowires, although the nature and localization of these channels is dependent on the specific distribution of charges around the nanowire. In such a scenario, the resonant feature in Fig. \ref{fig:Fig4} can be attributed to a defect or, more generically, a potential minimum localized near the wave functions of subband 2, but far from the wave functions of subband 1.
 
In conclusion, we presented conductance measurements reflecting the 1D band structure of an InSb nanowire under an in-plane magnetic field. By varying the potential landscape through the use of three bottom gates, we demonstrated large tunability of the inter-subband orbital spacing and, in particular, the unexpected possibility of robust, ground-state level degeneracies emerging at large $B$.
These findings bare relevance to the study of topological phases in 1D superconductor-semiconductor structures. The occurrence of such exotic phases is in fact subordinated to special requirements on the 1D subband structure and filling condition.
Finally, we also showed that discrete bound states can produce conductance resonances preceding the conductance plateaus. Whether such resonant bound states could account for the reported signatures of helical gaps in Ref. \cite{Kammhuber2016} is unclear, and we believe this possibility should deserve further investigation. The $B$-induced transverse localization discussed in the present work is expected to be a prominent effect in experiments where $B$ exceeds $\sim$1 T. At this field, the transverse localization, set by the magnetic length ($\approx 25$ nm), becomes smaller than the typical nanowire radius.

\section{Supporting Information}

Contains the method for extracting the lever-arm parameters, g-factors, and subband splittings; measurements of an additional resonance before the first conductance plateau; and calculations of the subband structure of the nanowire under external magnetic and electric fields.

\section{Acknowledgments}

We acknowledge financial support from the Agence Nationale de la Recherche, through the
TOPONANO project, from the EU through the ERC grant No. 280043. The calculations were run on the Marconi/KNL machine at CINECA thanks to a PRACE allocation. 
The authors declare no competing financial
interest.

\bibliography{Bibliography}

\end{document}